\documentstyle[prabib,aps]{revtex}

\begin{document}
\draft
\title{ Stability of Liquid Crystalline Macroemulsions }

\author{E.M. Terentjev}

\address{Cavendish Laboratory, University of Cambridge,
Madingley Road, Cambridge, CB3 0HE, U.K. }

\maketitle
\vspace{0.2cm}

\begin{abstract}
The thermodynamic stability of emulsions of liquid crystal
in water (glycerol) matrices is demonstrated for a wide range of
materials and concentrations. Coalescence is prevented by an
energy barrier for a topological ring defect formation in a
neck between the two merging droplets. There is a characteristic
size of emulsion droplets, typically tens of microns or more,
controlled by the balance of elastic and anchoring energies of
the liquid crystal. On removal of liquid crystallinity (by
raising the temperature above $T_{ni}$ in thermotropic nematic
materials, for example) the energy barriers for coalescence
disappear and emulsion droplets can merge quickly, controlled
only by the traditional kinetic effects. Practical applications
of this effect, as well as some wider theoretical implications
are discussed in the end.
\end{abstract}

\vspace{0.3cm}



Physics and chemistry of colloids and emulsions is rapidly
becoming, from an empirical domain of paint design and food
technology, an important area of fundamental research. The reason
for this is the recognition of the possibilities
and benefits of new applications generated from a fundamental
studies and, on the other hand,  more clear, challenging
and wide-ranging problems are seen in this area (a good example
being, for instance, the physics of membranes, which made an
impressive progress in the recent years \cite{memb}).

When two immiscible fluids are thoroughly blended together, an
emulsion is formed (or a colloid suspension, when the
characteristic viscosities of the fluids differ significantly).
The best known example are, of course, oil-in-water or
water-in-oil emulsions. There is an understanding, quite general
for colloids and emulsions, that their structural stability is a
kinetic concept and not a thermodynamic one \cite{bibette}. Some
emulsions have only a short lifetime before complete phase
separation, whereas others remain kinetically stable for years. In
order to prepare a stable emulsion there must be a surface active
material present  to protect the newly formed droplets from
immediate recoalescence. Such surfactant molecules, typically,
have two distinct parts --- hydrophobic and hydrophilic for
oil-water interfaces. By aggregating on such interfaces they also
reduce the surface tension and, when such a reduction is complete,
a microemulsion, {\it i.e.} the solution of microscopic micelles is
formed.  That represents a truly thermodynamically stable phase,
as opposed to the kinetic stability of macroscopic droplets. The
argument is simple: if the local energy of the interface covered
with surfactant is very small (``zero'' for practical purposes),
the entropy preference of having many small particles in favour of
the few big ones makes the molecular micelles a ground state of the
total free energy. If, on the other hand, the local interface
energy is considerable, the configurational entropy is irrelevant
and the minimal surface area of the completely separated phases is
the ground state --- which the system may take some time to reach,
however.  It is not easy to give a reference to such a broad and
well-studied area, but an interested reader, new to the field,
could find useful  information in \cite{clint,foods}.

The situation may change if an additional physical field
comes into the problem, bringing along its own energy and entropy
contributions. The simple example studied in this work is the
orientational order when one of the two emulsified fluids is a
liquid crystal. This question has been recently examined
theoretically, when a nematic liquid crystal is the majority
phase with surfactant-covered fluctuating interfaces inducing bulk
curvature deformations of the director \cite{ourmemb}. Here I
shall address the opposite end of the problem, when the
isotropic majority phase (water, for instance) has small droplets
of nematic liquid crystal dispersed in it. Surfactant is needed,
as usual, to reduce the interface energy and also to impose
strong anchoring boundary condition for the nematic director. The
basic result of this work is that there is an additional
contribution, the Frank elastic energy of the nematic, in the
analysis of emulsion stability. This energy creates a significant
barrier for the droplet coalescence and leads to a new effect
--- thermodynamic stability of macroemulsions.

Droplets of nematic liquid crystal in an isotropic fluid and the
related topological defect structures  have been extensively
studied before \cite{oleg}. Since the development of
polymer-dispersed liquid crystals has been formulated \cite{pdlc},
much more work have been done in this area \cite{others}.
However, in all cases the properties and effects of individual
droplets have been addressed, with authors looking at the
structure,  and field and temperature - driven transitions
\cite{we2}. Interesting physical effects here stem from the
topological constraint, imposed on the director field in a closed
volume by the anchoring condition on its surface, leading to
topological defects of total point charge {\small (+1)} for a
nematic and more complex structures in other phases.

In this paper I study the question of cooperative effects
for such droplets in a (dense) emulsion, focusing on the
mechanism of their coalescence, which one would expect to happen
by analogy with isotropic macroemulsions. First I explain the
experimental procedure of forming such an emulsion (certainly
well-understood by the authors of quoted papers) and the facts
about the coalescence kinetics. After that I propose a model
structure of the ``neck'' between the two merging droplets and
calculate the related energy barrier, which explains the
observations. In the end I shall briefly discuss practical
implications of this effect and also its relation to the wider
theoretical problem of the effect of topological defects on phase
ordering in various field theories.

\underline{\sc Experiment} --- Since we are modelling
oil-in-water emulsions, with the oil phase being a liquid crystal,
the following general rule for the choice of materials should
apply.  If the majority isotropic phase is water (or glycerol,
similar in properties, or their mixture), that is a fluid with
large molecular dipoles, then the ``oil'' liquid crystal should be
chosen with the least possible dipole. In this way the surfactant
molecules will have a clear preference for positioning their
hydrophilic and hydrophobic parts on the interface. With the above
choice of non-polar rod-like mesogenic molecules, the best
surfactant type would be an ionic one, with one long aliphatic
hydrophobic tail similar to the typical end group of a nematic
molecule.  Along these guidelines the weakly polar mesogenic MBBA
(Methoxybenzylidene-butilaniline, supplied by Aldrich) must be
given preference to, for instance, cyanobyphenils (another typical
nematic material,  but with significant molecular dipole in the
$C$-$N$ group).\footnote{In fact, MBBA also has a disadvantage,
since it is weakly miscible with water, which causes the eventual
loss of its mesogenic power. However, in a good
surfactant-mediated emulsion this mixing takes a very long time,
sufficient to allow  unambiguous observations.}   A simplest
choice of surfactant leads to CTAB (Cetyltrimethylammonium
bromide, supplied by Aldrich), which has a strongly polar
hydrophilic head and a single hydrophobic tail with 14-16 $CH_2$
groups. One expects this tail to align itself along the rod-like
mesogenic molecule and, in fact, CTAB is one of the best materials
used to treat liquid crystal cells for homeotropic anchoring.

The preparation of an emulsion is fairly standard. One dissolves
a small amount of CTAB in water, $3-5$ weight \%, taking
care to avoid the surfactant aggregation into various lyotropic
phases above the CMC (critical micelle concentration). Then a
small amount (also $\sim 5$ weight \%) of MBBA is added and the
resulting mixture is thoroughly blended in an ultrasonic bath. As
will be clear below, it is important to keep the ``oil'' all
the time in the nematic phase, below the clearing temperature
$T_{ni}$.  As a result, a very small sub-micron size emulsion of
MBBA in water, with CTAB aggregated on the interfaces, is formed.
Due to the initial small size of droplets the material scatters
light and appears turbid. After some time these droplets recombine
into much bigger ones ($\sim 30-100$ microns) and then the
dynamics stops: samples then stayed unchanged for many months.
The time of initial recombination can vary between minutes and
hours depending on conditions and concentrations, but the physics
and the resulting structure are fairly robust. For instance, since
the density of MBBA is
 slightly higher than of water ($d=1.027$), the droplets
eventually sink to the bottom of container. In order to improve
their buoyancy one may choose to mix water with glycerol
($d=1.26$) to equilibrate the density with MBBA. One also may
significantly alter the concentration of CTAB in the mixture, or
blend again at some moderate ultrasonic energy input. None of
these factors appear to qualitatively influence the resulting
stable macroemulsion.

In order to test the ``strength'' of macroscopic droplets in the
stable emulsion one can use pure water with its low viscosity
and density in comparison with the MBBA liquid crystal. (There
is no need to prepare a new sample, one can, at will, add or
remove water and surfactant from the clear regions of the
established emulsion, without any effect on droplet distribution).
All droplets quickly sink to the bottom, but can be disturbed by
shaking the container. In the stationary state the droplets are so
tightly compressed together in the pile by gravity, that their
shape no longer remains spherical, see Fig.1. And yet, in spite of
the physical force pushing droplets together, no coalescence takes
place over the observation time of several months. This fact is a
surprise, considering what we know about isotropic
oil-in-water emulsions. To test this comparison, the obvious
thing to do is to raise the temperature of the thermotropic
liquid crystal above its clearing point, after which the system
must behave like a ``normal'' o-w emulsion without an additional
elastic energy of the nematic.  On doing so a remarkable thing
happens: all droplets that were in direct contact (like those
shown in Fig.1) coalesce immediately, within a few seconds after
$T_{ni}$ is exceeded! This observation clearly proves that the
nematic curvature elastic energy holds the macroemulsion
stable.

\underline{\sc Theory} ---
How can we rationalize these observations? We have the following
facts: in the initial stage very small droplets recombine quickly
to form a stable macroemulsion with characteristic size $\sim 50$
microns (broadly distributed within this order of magnitude);
this size and droplet lifetime appears to be independent of the
character of the isotropic majority phase (as long as it supports
the CTAB placement on the interface); the application of
significant force pushing droplets against each other is not
sufficient to make them merge and, finally, the coalescence and
complete phase separation take place very quickly after the liquid
crystallinity is removed from the picture.

All following arguments will be based on the assumption that
hydrophobic tails of densely packed surfactant (CTAB in this
case) are extended perpendicular to the interface and impose a
strong anchoring on the nematic director. (We certainly know this
is the case with CTAB-treated glass surfaces). The deviation
$\delta  {\bf n}$ of the director from this perpendicular
orientation is penalized by the anchoring energy  $ \sim W
|\delta {\bf n}|^2$ per unit area, while the curvature
deformations in the bulk give rise to the Frank energy density
$\frac{1}{2}K (\nabla \delta {\bf n})^2$. Big droplets with the
radius $R^* \geq K/W$ form a radial hedgehog point defect in the
middle due to the topological constraint at non-zero $W$ (or,
possibly, a small ring of disclination \cite{mori}, which would be
undistinguishable from the monopole at large length scales); this
is what we see in Fig.1 and in either droplet of Fig.2.

Let us consider the act of two droplets coalescence in some more
detail, see Fig.2. As separate bounded volumes before collision,
each of them contains a topological point charge of {\small (+1)}
(regarding the director as pointing outwards from the monopole).
After the merging process is completed we are left with one
piece of bounded nematic volume, which has just one {\small (+1)}
point defect. Hence, in between, another defect with point charge
{\small (-1)} was born and then annihilated one of the two initial
monopoles. This defect is, as one can see from the Fig.2, a ring
of the wedge disclination with the {\it linear} charge {\small
(-1/2)}  on the tip of the neck connecting the two droplets. The
late-stage scenario can be different: one possibility is that the
ring detaches itself from the surface, when the curvature of the
neck decreases, and drifts into the volume towards one of the
radial monopoles, the other is that the {\small (-1/2)} ring
remains as a surface disclination \cite{oleg} and one of the point
defects is, being attracted to it, moves towards and disappears on
the surface. This choice is curious but is of little relevance to
the practical phenomena; what is important is that in order to
form the initial neck the system must create an accompanying
total {\small (-1)} point charge not as a {\it pair} (a
fluctuational birth of small {\small ($\pm $1)} pairs should be a
normal event in a vector-field system), but in isolation, at a
distance $\sim R$ from other defects. This costs a macroscopic
elastic energy $\sim KR$.

Now we can trace the evolution of the nematic emulsion. Initial
droplets, broken to very small sizes by ultrasonic energy,
prefer to violate the surface anchoring energy of the director
(at a cost $\sim WR^2$) but have a uniform director field inside.
No topological defect arguments apply in this situation and small
droplets recombine without an energy barrier. At a critical size
$R^* = K/W$ the surface energy cost becomes too high and droplets
form a radial director distribution with a monopole in the
middle, with bulk elastic energy $\sim KR$. From now on
each individual pair of droplets, in order to merge, would have
to overcome the topological energy barrier $\Delta E \sim KR \sim
K^2/W$. We can now understand why the conditions {\it outside}
the droplet do not affect the emulsion stability, as long as the
strong anchoring constraint imposed by the ordered surfactant is
preserved. Near the nematic--isotropic transition the Frank
elastic constant $K$ and the anchoring energy $W$ depend on the
order parameter $Q$, as $K \sim Q^2 $ and $W \sim Q$ and,
therefore,  the energy barrier to coalescence $\Delta E \sim Q^3$
rapidly disappears near the transition.

The same estimate, $\Delta E \sim K^2/W$, controls the parameter
of another test on the emulsion stability. We saw already that
the weak force of gravity is not sufficient to overcome this
barrier. However, if we place the established emulsion in an
ultrasonic field, the droplets may again break into very small
sizes. Note, however, that in order to break the droplet in two
parts the same topological analysis demands the
formation of a disclination ring on the surface, preferably in
the breaking-neck region, in order to provide {\it two} point
singularities after the break. The energy estimate is the same,
$\Delta E \sim KR$, and dictates the threshold value for the
ultrasonic energy input of a blender (shaking by hand is certainly
not enough to provide this energy per droplet). For a typical
nematic liquid crystal the Frank constant is of the order $K \sim
10^{-11} \, \hbox{N}$ and ``strong'' surface anchoring
usually means $W \sim 10^{-5}-10^{-6} \, \hbox{J/m}^2$. From this
we obtain the correct characteristic stable droplet size $R^*
\sim 10-100$ microns and the energy barrier per particle
$\Delta E \sim 10^{-16} \, \hbox{J}$ (compare with
room-temperature thermal energy $k_BT \sim 5 \, 10^{-21} \,
\hbox{J}$). Note that at weak anchoring the characteristic size
$R^*$ would increase and the spherical droplets may transform
into large flexible vesicles (in practice this happens at $R \sim
1-10 \hbox{mm}$), after which the energy barrier $\Delta E$ will
decrease again. Thus, when the director anchoring on
surfactant-treated interfaces is weak, no stable emulsion should
be expected.

\underline{\sc Conclusions} ---
Besides the possibility of thermodynamic stability of
macroemulsions, the main practical implication of this work is
the ability to have a fine temperature control over the
morphology of emulsions. Also of interest is the control over the
equilibrium droplet size $R^*$ through the choice of surfactant
(affecting $W$) and nematic material. One can imagine, among many
other possibilities, an application of such emulsions, carrying a
liquid crystalline chemical inhibitor, temperature-tuned in such
a way that emulsion becomes isotropic, rapidly coalesces and
releases the chemical to stop an overheating reactor. Another
possible application could use the polymerization of surfactant
to form a percolating sponge with mesh size controlled by $K/W$.

Another, more fundamental aspect of this work is related to the
general problem of phase ordering of a non-scalar field. Such a
problem arises in various areas of physics, from cosmology to
soft condensed matter, where a first order ordering phase
transition takes place. Currently  general opinion is in
favour of the so-called Kibble mechanism, when {\it uniform}
nucleating domains merge and form topological defects in the
ordered phase due to their random orientation \cite{cosmo}. There
are various opinions on why point defects (monopoles) are never
observed after such quenching. However, there seems to be no good
reason to disregard the additional symmetry breaking due to the
order parameter gradient on the nuclei surface and corresponding
surface energy (equivalent to our $W$). Then, if the
characteristic size $R^*$ were to be reached, the domains would
be topologically non-trivial and their coalescence will follow the
route shown in Fig.2, completely altering the predictions about
the broken-symmetry phase properties after the quench.
\medskip

I gratefully acknowledge many useful and enlightening discussions
with J.A. McDonald,  R. Peck, O.D. Lavrentovich, P.
Palffy-Muhoray, M. Warner and P.D. Olmsted .  This research has
been initiated and supported by Unilever PLC.

\newpage

\figure{ FIG.1 \ \ A group of tightly squeezed nematic droplets
in water/glycerol/CTAB matrix, viewed between crossed polarizers
to reveal the radial director distribution. The bar represents
the $50$ micron length scale. Note the squeezed droplets in the
bottom-left corner of the picture. \label{droplets}}

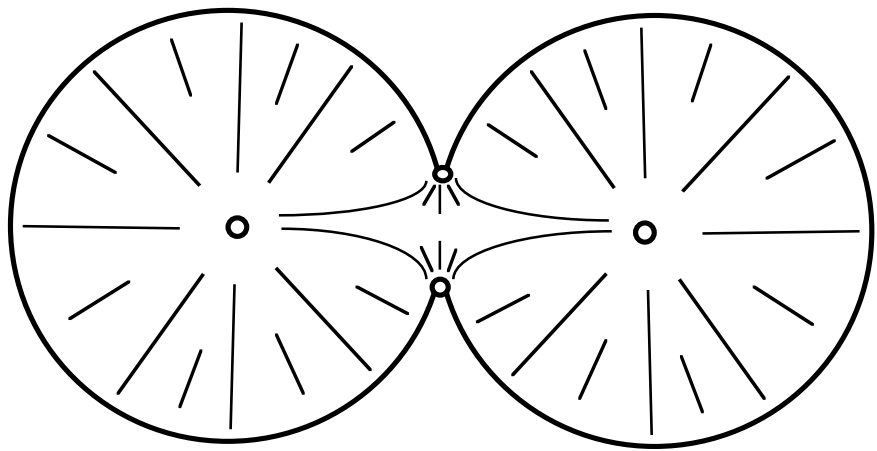
\figure{ FIG.2 \ \ A scheme of the neck formation in the
coalescing spherical droplets (the picture represents a
cross-section of a cylindrically symmetric dumb-bell object). The
resulting topological ring-defect, situated on the neck tip, is
required to conserve the total topological charge of the merged
system. \label{neck}}


\begin{references}
%
\bibitem{memb}
W. Helfrich, {\it J.~Phys (France)} {\bf 46}, 1263 (1985); \
R. Lipowsky and S. Leibler, {\it Phys. Rev. Lett.} {\bf 56},
2541 (1986); \  D. Roux, {\it  Physica} {\bf A172}, 242 (1991).
%
\bibitem{bibette} J. Bibette, D.C. Morse, T.A. Witten and D.A.
Weitz, {\it Phys. Rev. Lett.} {\bf 69}, 2439 (1992).
%
\bibitem{clint} J.H. Clint, {\it Surfactant Aggregation},
Blackie, Glasgow -- London, 1992.
%
\bibitem{foods} E. Dickinson, {\it An Introduction to Food
Colloids}, Oxford University Press, 1992.
%
\bibitem{ourmemb} P.D. Olmsted and E.M. Terentjev, to be
published in {\it J. Physique II}.
%
\bibitem{oleg} G.E. Volovik and O.D. Lavrentovich, {\it Sov. Phys.
JETP}, {\bf 58}, 1159 (1983).
%
\bibitem{pdlc} S. Zumer, A. Golemme and J.W. Doanne, {\it J.
Optical Soc. A}, {\bf 6}, 403 (1989).
%
\bibitem{others} F. Xu, H.S. Kitzerow and P.P. Crooker, {\it
Phys. Rev. A}, {\bf 46}, 6535  (1992); \ P.S. Drzaic and A.
Muller, {\it Liq. Cryst.}, {\bf 5}, 1467 (1989).
%
\bibitem{we2} O.D. Lavrentovich and E.M. Terentjev, {\it Sov.
Phys. JETP}, {\bf 64}, 1237 (1986);
\ A.E. Koehler, {\it Z. Phys. Chem.}, {\bf 269}, 196 (1988); \
S. Kralj and S. Zumer, {\it Phys. Rev. A}, {\bf 45}, 2461 (1992).
%
\bibitem{mori}H. Mori and H. Nakanishi, {\it J. Phys. Soc. Japan}
{\bf 57}, 1281 (1988).
%
\bibitem{cosmo} M.B. Hindmarsh and T.W.B. Kibble, {\it
Rep. Prog. Phys.} {\bf 58}, 477 (1995); \ A.J. Bray, {\it Physica
A} {\bf194}, 41 (1993).
%
\end{references}
\end{document}